\documentclass[aps,pre,showpacs,twocolumn,superscriptaddress]{revtex4}
\usepackage{graphicx}
\usepackage{amsmath}
\usepackage{amssymb}
\bibstyle{apsrev.bib}

\newcommand{\be}{\begin{equation}}
\newcommand{\ee}{\end{equation}}
\newcommand{\bea}{\begin{eqnarray}}
\newcommand{\eea}{\end{eqnarray}}

\begin{document}

\title{Subgraph Ensembles and Motif Discovery Using a New Heuristic for Graph Isomorphism}
\author{Kim Baskerville}
\affiliation{Perimeter Institute for Theoretical 
Physics, Waterloo, Canada, N2L 2Y5}
\affiliation{Complexity Science Group, Department of Physics and Astronomy,
University of Calgary, Calgary, Alberta, Canada T2N 1N4} 
\author{Maya Paczuski}
\affiliation{Complexity Science Group, Department of Physics and Astronomy,
University of Calgary, Calgary, Alberta, Canada  T2N 1N4}

\date{\today}

\begin{abstract}
  A new heuristic based on vertex invariants is developed to rapidly
  distinguish non-isomorphic graphs to a desired level of accuracy.
  The method is applied to sample subgraphs from an E.coli protein
  interaction network, and as a probe for discovery of extended motifs.
  The network's structure is described using
  statistical properties of its $N$-node subgraphs for $N\leq 14$. 
  The Zipf plots for subgraph occurrences are
  robust power laws that do not change when rewiring
  the network while fixing the degree sequence --- although the
  specific subgraphs may exchange ranks. However the exponent depends
  on $N$. The study of larger subgraphs highlights some striking
  patterns for various $N$. Motifs, or connected pieces that are
  over-abundant in the ensemble of subgraphs, have more edges, for a
  given number of nodes, than antimotifs and generally display a
  bipartite structure or tend towards a complete graph. In contrast,
  antimotifs, which are under-abundant connected pieces, are mostly
  trees or contain at most a single, small loop. The extension to
  directed graphs is straightforward.
\end{abstract}
\maketitle
\section{Introduction}
\label{sec:intro}

The recent surge of interest in complex networks has often targeted
general features of organisation
\cite{gp1,gp2,gp3,gp4,gp5,gp6,gp7,gp8,gp9,gp10}. A number of common
properties have been observed, including the so-called small world
effect, fat tails in the distribution of the node degree (the
``scale-free'' network), as well as clustering.  Although the last two
attributes are statistical properties of the local network structure,
networks that share these features may nonetheless exhibit totally
different specific local structures.  Certain connected subgraphs with
three or four nodes, termed ``motifs'' \cite{mot1,mot2,mot3}, turn out
to be significantly over-abundant in real networks when compared to
null models.  These null models are typically randomised networks
where the smaller scale structure (e.g. node degree)~\cite{ms} is
determined by the original network. It is believed that networks with
similar functions -- for example, forward logic chips and neural
networks -- display the same motifs~\cite{mot1}. A growing body of
evidence indicates that particular motifs perform specific functions
in gene transcription networks
\cite{mot2,func1,func2,func3,func4,func5,func6,func7,func8}. In
addition, proteins within motifs are more conserved across species
than proteins that do not form part of such units
\cite{conserve1,conserve2}.

Motifs and antimotifs, which are significantly under-abundant
connected subgraphs, may also be useful in classifying networks and
comparing real-life situations to theoretical models. Milo {\it et
  al.}  \cite{sp} explored significance profiles: normalised
$Z$-scores for particular connected subgraphs.  They claim to find
``superfamilies'' of networks displaying similar profiles. In a
similar vein, Middendorf {\it et al.}~\cite{class1} used exhaustive
subgraph enumeration of networks generated by different theoretical
models as training data for a machine learning algorithm, and
developed a discriminative classifier subsequently able to identify
new networks with success.

However, all of these approaches have been handicapped by the small
size of connected subgraphs. This limits the scale where features of
organisation in networks can be discovered.  In most cases, connected
subgraphs with at most four nodes are considered.  Middendorf {\it et
  al.}~\cite{class1} searched for two different categories of
subgraphs: graphs which could be generated by a random walk of length
less than or equal to eight, and graphs with up to seven links - to
achieve slightly larger subgraphs.  Ziv {\it et al.}~\cite{ind2}
analysed statistically significant measures that can be calculated
directly from the adjacency matrix. These measures are related to
subgraphs but lack a one-to-one correspondence.  Hence the possibility
of insight into the function of organised structure at different
scales or the systematic discovery of larger scale structures is -
from our point of view - lost.

The existing size limitation for motif discovery leaves some
interesting questions unanswered. Do motifs appear independently, or
do they combine to form larger organised
structures~\cite{ind1,ind2,ind3} that are overwhelmingly represented
in the real network compared to an appropriate null model?  If so,
what do these extended structures look like? What properties of the
network's ensemble of $N$-node subgraphs distinguish it from null
models or from other networks?  Are collections of nodes that
participate in motifs of larger sizes also more likely to be related
to function and/or conserved through evolutionary history? Kashtan
{\it et al} made some progress in this direction by considering
specific generalisations of three and four node motifs~\cite{gen1}.
They found that networks sharing a particular three node motif
favoured different generalisations of that motif, suggesting that
larger structures need to be considered to fully understand how the
network is organised.  However, this work relied on {\it a priori}
assumptions about possible generalisations to larger motifs.  Searches
were tailored to particular subgraphs. A more general analysis is
known to be computationally difficult~\cite{samp,comp1,comp2,top1}.

\subsection{Problems in Finding Extended Structures}

There are at least three main problems. The first is that the time
required for exhaustive enumeration of subgraphs increases rapidly
with subgraph size, particularly for large networks. This can be
solved by sampling: Kashtan et al \cite{samp} showed that quite small
samples could be sufficient to identify motifs with up to seven nodes.
However, their method requires the calculation of weights in order to
achieve uniform sampling.  Their calculation of these weights
increases in difficulty, with combinatorial factors, as the the
connected subgraph size increases.  We achieve uniform sampling
automatically by picking nodes at random from the network -- at the
expense of sampling both connected and disconnected subgraphs.

The second problem is to determine appropriate null model(s) and
significance. The standard null model (see for instance
Ref.~\cite{ms}) is where the degree of every node is not allowed to
change -- so the single node properties are fixed.  Such an ensemble
can be obtained using a Sequential Monte Carlo method called
"rewiring". Statistically significant deviations from that background
are by definition coming from node-node correlations.  Extending this
argument, when Milo {\it et al.}~\cite{mot1} search for 4-node motifs
they also fix the actual number of each kind of 3-node subgraph in
their null model.  However, as in Ref.~\cite{gen1}, here we use only
the ensemble of fixed degree sequence as a null model to test for
significance. Explicitly fixing the occurrence of $(N-1)$-node
subgraphs is computationally intractable for larger $N$. There are not
only linear constraints between different subgraphs arising from
conservation laws (see Ref.~\cite{sp} and Section~\ref{sec:linear})
associated with rewiring but also non-linear correlations caused, in
part, by the form of the null model.

The third difficulty lies in distinguishing non-isomorphic subgraphs.
This is the well-known and notoriously difficult ``graph isomorphism
problem'' \cite{graph1,graph2}. The number of possible graphs grows
faster than exponentially with $N$~\cite{top1}. Several
algorithms~\cite{prog1,prog2,prog3,prog4,prog5} are available, but
most of these are configured to make a comparison for isomorphism
between two graphs. Comparing each new subgraph pairwise to all
subgraphs already identified would be far too time-consuming in this
context.  Some existing programs can be altered to provide sets of
labels to identify particular graphs.  They tend to be optimised for
large graphs (hundreds of nodes), and appear to us to be unsuitable
for the type of search required for discovery of organisation at
larger scales than three or four nodes.

At this point in time, discovery of larger scale organisation does not
require particularly large subgraphs.  Ten to fifteen nodes would
already be a significant step forward, and entails a new set of
problems and types of behaviours as discussed later.  Subgraphs do,
however, need to be classified quickly if a method is to be practical.
We present a new heuristic that assigns a set of labels to each
subgraph as it is sampled, so that isomorphic graphs are guaranteed to
have the same label(s), but (most) non-isomorphic graphs have
different labels. The accuracy of the method depends on the number of
labels used -- at the expense of increased computational effort.  We
test the heuristic by comparing with exact enumeration of all
isomorphic graphs up to $N=8$. Combined with a sampling technique, our
heuristic is used to identify extended motifs of a protein
interaction network. We sample both connected and disconnected
subgraphs uniformly by picking $N$ distinct nodes at random.  Motifs
are then discovered by looking at the significance -- with some
caveats -- of individual subgraphs that contain these structures as
distinct pieces.

\subsection{Summary}

The labelling algorithm is described in Section~\ref{sec:algorithm}.
In Section~\ref{sec:evaluation} various stages of the algorithm are
tested. The full algorithm successfully distinguishes all graphs with
up to eight nodes~\cite{foot1}. Differences in the running times and
accuracy of the stages are also discussed. In
Section~\ref{sec:motifs}, the algorithm is used to identify extended
motifs and antimotifs in the E. coli protein interaction network. The
motifs all share a remarkably similar bipartite structure, which is
completely different from the long chains and tails seen in
antimotifs. For fixed $N$ the distribution of all subgraph counts is found to
obey a power law, where the exponent depends on $N$. However, the Zipf
plots of the real and randomised networks are quite similar although
the subgraphs exchange rank. In Section~\ref{sec:conclusions} we
conclude with a summary.

\section{The labelling algorithm}
\label{sec:algorithm}

The algorithm developed here can be applied to both simple graphs and
digraphs -- graphs with directed edges. Here we will concentrate on
the algorithm for simple graphs, leaving the straightforward
generalisation to digraphs to a later publication.

Motif discovery requires a fast way to identify graphs that are
isomorphic.  One way to be certain that two graphs are isomorphic is
to find the isomorphism that maps one to the other.  This is a
permutation of the vertex labels of one graph such that its list of
links becomes identical to that of the other graph. To show that two
graphs are not isomorphic therefore requires proving that no such
isomorphism exists, which in theory requires checking every possible
permutation of the vertices. Since there are $N!$ such permutations
for a graph with $N$ nodes, this is far too time-consuming to be
practical. Many algorithms therefore start by trying to reduce the
number of permutations that need to be checked, usually by applying
some kind of ``canonical labelling''~\cite{gp2} or ordering to the
vertices. For example, if a unique way of ordering the vertices in
both graphs can be found, then vertices of the same rank must map to
each other -- in order for the graphs to be isomorphic.

\begin{figure}[!tb]
\includegraphics[angle=0,width=8.0cm]{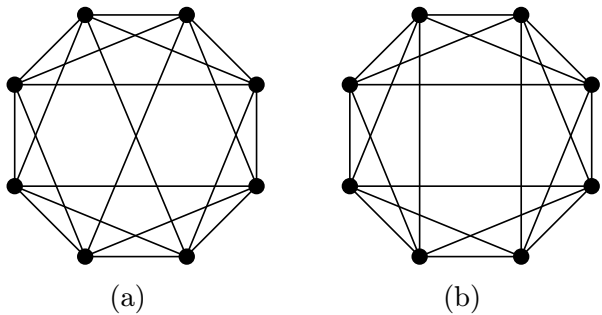}
\caption{Two non-isomorphic graphs that cannot be distinguished by any of the
  invariants proposed by Remie.}
\label{fig:BPgraphs}
\end{figure}

An alternative approach is to try to find an invariant under
permutation, or set of invariants, that uniquely labels any graph. The
use of invariants ensures that isomorphic graphs always receive
identical labels.  However it is not certain that non-isomorphic
graphs will receive at least one different label.  Remie~\cite{remie}
defines four different invariants, but none of these can distinguish
the eight node graphs in Fig.~\ref{fig:BPgraphs} as non-isomorphic.

\subsection{Invariant Vertex Labels}

Our approach defines vertex invariants through a generalisation of
standard canonical labelling~\cite{gp2}.  Usually, the canonical label
depends only on the degrees of the vertex being labelled together with
its immediate neighbours. This means, for example, that all vertices
in a long chain (except the two endpoints) receive the same label,
whereas it is clear that nodes near the end of the chain should be
distinguishable from nodes nearer the middle. Bearing this in mind, we
have extended the usual canonical labelling to include all vertices in
the graph. In the case of a graph made of disconnected pieces, we
include all vertices in the connected piece containing the vertex being labelled.

As with usual canonical labelling, our label is a sum of powers of
two, with the vertex degrees, ${k_j}$, determining the power. To
include all vertices, but give a higher weight to those closest to the
vertex $V_{i}$ being labelled, we include an additional factor of
$2^{x-x_{ij}}$, where $x$ is the diameter. This diameter is the
maximum shortest path between any two vertices on the connected piece
of the subgraph containing $V_{i}$.  The quantity $x_{ij}$ is the
distance between vertices $V_{i}$ and $V_{j}$, where $V_j$ is required to 
be connected to $V_i$ by some path. The lowest possible
weighting is $2^{0}=1$ (if $x_{ij}=x$), and the highest weighting
($2^{x}$) is given to $V_{i}$ itself.  Each vertex
$V_{i}$ is assigned a label $X_{i}$ as follows:
\be
\label{eq:xi}
X_{i} = \sum_{j}^{\rm connected} 2^{x-x_{ij}+ k_{j}} \quad ,
\ee 
where $k_j$ is the degree of vertex $V_{j}$ \cite{bases}.  The sum is
taken over all vertices in the graph, or if the graph contains several
disjoint subgraphs, over all vertices in the connected subgraph
containing $V_{i}$.

\begin{figure}[!tb]
\includegraphics[angle=0,width=8.0cm]{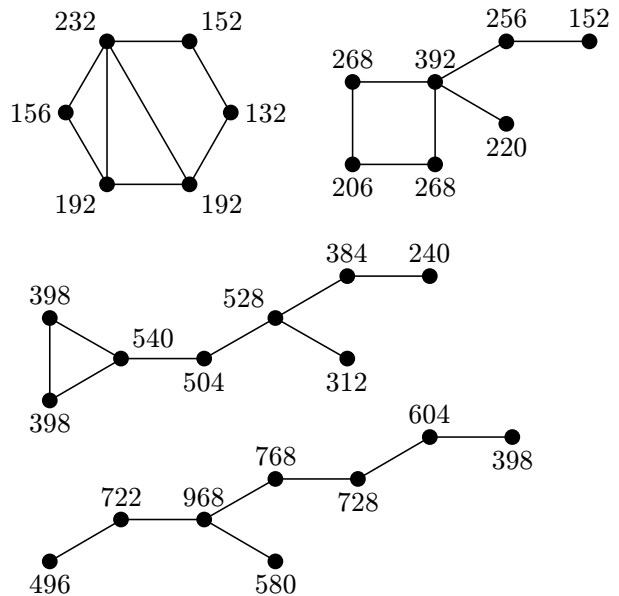}
\caption{Vertex labels calculated using Eq.~(\ref{eq:xi}). Higher
  values are assigned to more central vertices, or those with higher
  degrees.}
\label{fig:examples}
\end{figure}

The labels defined by Eq.~\ref{eq:xi} have an intuitive meaning.  More
connected or central vertices have higher values.
Fig.~\ref{fig:examples} gives some examples of the labelling scheme
for different subgraphs. The labels $X_{i}$ are clearly higher for
more central vertices than those closer to the edge.

\subsection{Invariant Graph Labels}

\begin{figure}[!tb]
\includegraphics[angle=0,width=8.0cm]{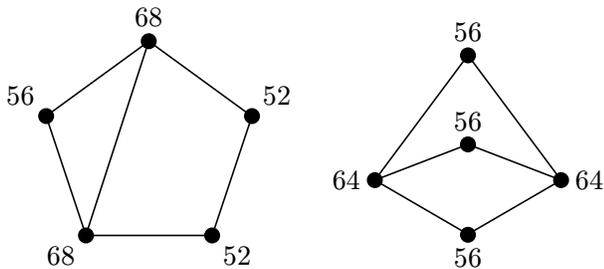}
\caption{These two graphs have different vertex labels $X_{i}$, which
  nonetheless combine to give the same sum:
  $l_1'=68+68+56+52+52=64+64+56+56+56=296$.  Their graph labels $l_1$,
  however, are not equal.}
\label{fig:n5problem}
\end{figure}

The set of vertex labels could be used directly to distinguish graphs,
but they would need to be ordered, for instance in descending order,
before comparisons between graphs could be made.  Another approach is
to combine the vertex labels to obtain a small set of graph labels.
One candidate graph label is the sum $l_1'=\bigl(\sum_i X_{i}\bigr)$.
Unfortunately it does not produce unique labels.
Fig.~\ref{fig:n5problem} shows two graphs that have the same sum
despite having different vertex labels (and hence being clearly
non-isomorphic).  However the product does not suffer from this
defect. In theory it could, but in practice we have not found it to be
the case for the graphs studied.  Our first graph label is therefore
defined to be
\be
\label{graphlabel1} 
l_{1} = \prod_{i} X_{i} \quad .
\ee 

Note that this product is over all the vertices in the graph, whether
it is connected or made of disjoint pieces.  Should this product
become too large to be conveniently stored as an integer, the first
several (eg.~9) digits can be used instead, without causing any
degeneracy in labels.  Again, this is an empirical observation rather
than a mathematical certainty. However, this is not the end of the
story.

We found that $l_1$ successfully distinguishes all graphs with up to
five nodes, but there are two pairs of non-isomorphic graphs with six
nodes that are assigned identical values. The graphs in
Fig.~\ref{fig:BPgraphs} provide another problematic example. These
graphs are highly symmetric.  In both graphs, every vertex has degree
five -- with the remaining nodes at distance $x_{j}=2$. Hence all the
labels, $ X_{i} = 2^2 * 2^5 + 5*2^1*2^5 + 2*2^0*2^5 = 576$, are
identical.

\begin{figure}[!tb]
\includegraphics[angle=0,width=8.0cm]{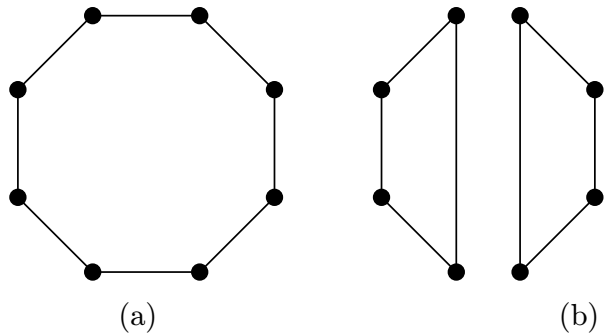}
\caption{These two graphs are the complements of the graphs shown in
  Fig.~\ref{fig:BPgraphs}.}
\label{fig:BPcomp}
\end{figure}

If all vertices are equally ``connected'', but the two graphs are not
isomorphic, what is the difference between them? Taking their
complements (exchanging links and non-links for every vertex pair) as
shown in Fig.~\ref{fig:BPcomp} reveals the source.  While the
complement of graph (a) is a single loop with 8 links, which we shall
now refer to as an 8-loop, that of graph (b) consists instead of two 4-loops.
Applying our labelling method to these complements produces unique
labels, which suggests a possible solution to the problem. For all
graphs, first calculate $l_{1}$ as described above.  Then take the
complement of each disconnected subgraph of the graph.
Recalculate labels, $Y_{i}$, for this new graph, and combine these
labels into the product
\be
\label{graphlabel2} l_{2} = \prod_{i} Y_{i} , 
\ee 
where the product is again taken over all vertices in the graph.  Each graph
is then labelled by the vector $(l_{1},l_{2},L,N)$, where $L$ is the
total number of links in the graph. Note that for disjoint graphs, it
is extremely important to \emph{take the complement of each connected
  subgraph individually}; if the complement of the whole graph is
taken instead, small disconnected pieces can cause problems, so that
degeneracy in labelling appears for quite small graphs.
An algorithm with these graph labels was tested by applying it to
every possible labelled graph for $N\leq 8$ and measuring the number
of distinct sets of labels. This number was compared to the true
number of non-isomorphic graphs.  Those were determined using Polya's
enumeration theorem. The algorithm uniquely labelled every graph with
up to six nodes ($N=6$), distinguished 1038 out of 1044 for $N=7$ and
12078 out of 12346 for $N=8$.  Even for $N=8$ almost $98\%$ of
distinct graphs were uniquely labelled.

What further invariant properties can be used as labels? Again,
considering the complements in Fig.~\ref{fig:BPcomp} provides a clue --
their different loop structures.  In fact the
numbers of all loops except 3-loops are different for the two graphs in
Fig.~\ref{fig:BPgraphs}.  We counted all the loops
in a graph by searching through its adjacency matrix. The number of
3-loops ($n_{3}$), 4-loops ($n_{4}$) etc.~can then be incorporated as
extra labels, so that each graph is labelled by the vector
$(l_1,l_2,L,n_3,n_4,...,n_N)$. This adapted algorithm, when tested,
correctly distinguished \emph{all} graphs with up to $N=8$ nodes.
Exhaustive testing of graphs with more nodes is not worthwhile at
present, as the program for $N=9$ would run for more than a year on a
present day standard laptop.

\section{Testing the algorithm}
\label{sec:evaluation}

This section may be skipped by those primarily interested in 
motif discovery.  As stated in Section~\ref{sec:algorithm}, all
stages of the algorithm have been tested exhaustively for graphs with
up to eight nodes. A simple graph with $N$ nodes contains $L_{MAX}={N
  \choose 2} = N(N-1)/2$ vertex pairs. Thus $L_{MAX}$ is the maximum
possible number of links, and $2^{L_{MAX}}$ is the number of labelled
graphs.  An easy way to generate all labelled graphs is to cycle
through the binary numbers between 0 and $2^{L_{MAX}} - 1$, loading
their digits in order into the off-diagonal elements of an adjacency
matrix. The labelling algorithm can then be successively applied to
each matrix or graph.  The accuracy of the algorithm can be evaluated
by comparing the number of graphs correctly distinguished to the true
number of non-isomorphic graphs, as determined by Polya's enumeration
theorem. The results for different stages of the algorithm are shown
in Table~\ref{tab:countgraph}. Note that since the labels are
invariants, isomorphic graphs must be assigned the same set of labels.
Thus it is not possible to overcount the number of distinct graphs.
Undercounting is possible, however, since non-isomorphic graphs may
nonetheless have similar enough structures to produce degenerate
labels.

\begin{table}[!b]
\begin{ruledtabular}
  \caption{Number of graphs distinguished by different graph labels
    compared to the exact number of graphs calculated using Polya's
    enumeration theorem, shown in the second column.  The third column
    shows the result obtained by using the sum, $l_1'$, rather than a
    product, $l_1$, of the vertex labels.  In the remaining columns
    $l_{1}$ and $l_{2}$ are as defined in
    Equations~(\ref{graphlabel1}) and~(\ref{graphlabel2}). The last
    column includes the number of loops as graph labels.}  {\vskip
    6pt}
\label{tab:countgraph}
\renewcommand{\arraystretch}{1.25}
\begin{tabular}{|c|c|c|c|c|c|}
\hline
 & \multicolumn{5}{c|}{\textbf{Number of Graphs}} \\
$\mathbf{N}$ &\textbf{Exact} & $\mathbf{l_{1}'}$ &
$\mathbf{l_{1}}$ & $\mathbf{l_{1}}$, $\mathbf{l_{2}}$ & 
 $\mathbf{l_{1}}$, $\mathbf{l_{2}}$, \\
 & & \textbf{(sum)} & & & \textbf{loops} \\ \hline
2 & 2 & 2 & 2 & 2 & 2 \\ \hline
3 & 4 & 4 & 4 & 4 & 4 \\ \hline
4 & 11 & 11 & 11 & 11 & 11 \\ \hline
5 & 34 & 33 & 34 & 34 & 34 \\ \hline
6 & 156 & 136 & 154 & 156 & 156 \\ \hline
7 & 1044 & 693 & 1004 & 1038 & 1044 \\ \hline
8 & 12346 & 4381 & 11188 & 12078 & 12346 \\ \hline
\end{tabular}
\end{ruledtabular}
\end{table}

\begin{figure}[!tb]
\includegraphics[angle=0,width=8.0cm]{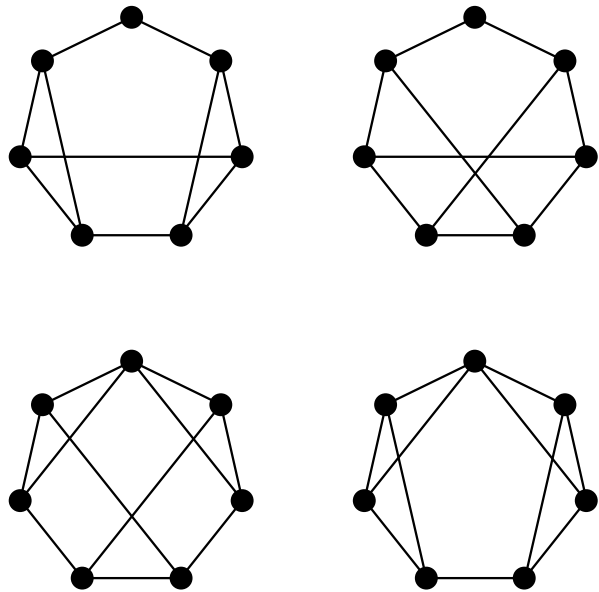}
\caption{The bottom pair of graphs are the complements of the top
  pair.  Neither pair can be distinguished by $l_{1}$ or $l_{2}$. The
  pairs exhibit different loop structures and can therefore be
 differentiated by loop enumeration.}
\label{n7missing}
\end{figure}

Table~\ref{tab:countgraph} shows that incorporating loop counting
together with $l_1$ and $l_2$ is the most accurate method. However the
cost in computing time is significant. On a standard laptop, for $N=8$
it took four and a half hours to compute $l_{1}$ alone, six hours to
compute $l_{1}$ and $l_{2}$, and twenty six hours for the full
algorithm including loop counting. Using $l_{1}$ and $l_{2}$ without
loop counting works perfectly up to $N=6$, but then misses 6 graphs
(0.6\%) at $N=7$ and 268 graphs (2.2\%) at $N=8$.  The graphs shown in
Fig.~\ref{n7missing} are typical examples of pairs not distinguished
by either $l_1$ or $l_2$.  One graph can be mapped to the other by switching
the endpoints or "rewiring"  two links.  The complements of the
graphs share the same property; hence the degeneracy in $l_{2}$ as
well as $l_{1}$.

\begin{figure}[!tb]
\includegraphics[angle=0,width=8.0cm]{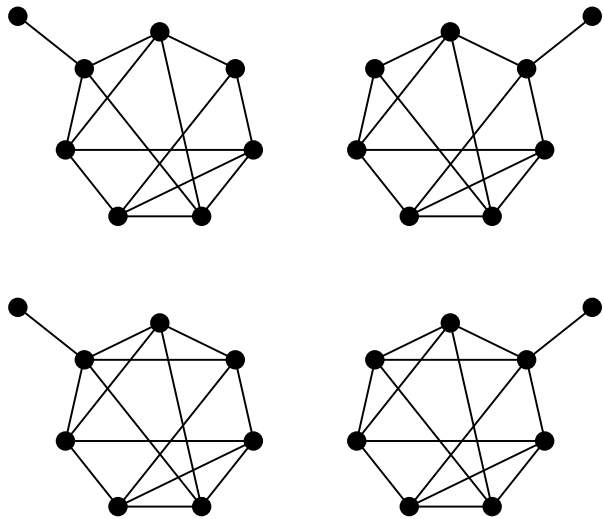}
\caption{The top and bottom pairs have the same vertex labels and the
  same loop structure. Both pairs are distinguished by the labels of
  their complements. Note that the bottom pair are identical to the
  top -- save for the addition of one extra link.}
\label{n8label2}
\end{figure}

Another possible route might be to omit $l_{2}$ when loop counting is
included. Using $l_{1}$ plus loop counting works perfectly up to
$N=7$, but fails to distinguish two pairs of graphs at $N=8$ (see
Fig.~\ref{n8label2}). The danger, as with omitting loop counting, is
that once an algorithm misses even a small percentage of graphs for
some $N$, it misses more and more as $N$ increases.

To summarise: The combination of $l_{1}$, $l_{2}$ and loop enumeration
differentiates all non-isomorphic graphs with up to eight nodes.
However, loop counting is very time consuming, and omitting it only
causes around $2\%$ of the $N=8$ graphs to be degenerately labelled.
With the above mentioned caveats we proceed with a subgraph census obtained by sampling
a protein interaction network using the algorithm with $l_1$ and
$l_2$, but without loops.

\section{Ensembles of Subgraphs and
Motif Detection in a Protein Interaction Network}
\label{sec:motifs}

We now present results for the statistics of subgraphs in the protein
interaction network of E. coli~\cite{ecoli}. The histogram of all
non-isomorphic subgraphs in the network is a characterisation of that
network.  This is termed a ``subgraph census''~\cite{wassermann_book}.
The ensemble of subgraphs is obtained by uniform sampling rather than
exact enumeration. This should give an accurate picture of the true
census up to statistical fluctuations and an overall normalisation.
Uniform sampling of connected and disconnected $N$-node subgraphs is
achieved by picking $N$ nodes at random.  Results were compared with
exact enumeration for small $N$.  Since there is no inherent
directionality in the interactions themselves, we have chosen to treat
the network as undirected. The network has 270 nodes and 716 links;
however it is not fully connected: seventeen pairs of nodes connect
only to each other, and there are two isolated triplets. The largest
connected component consists of 230 nodes and 695 links. Both this
piece, termed the giant component (GC), and the entire network are
studied.

\subsection{Zipf's Law for Subgraph Census}

\begin{figure}[!tb]
\includegraphics[angle=0,width=8.0cm]{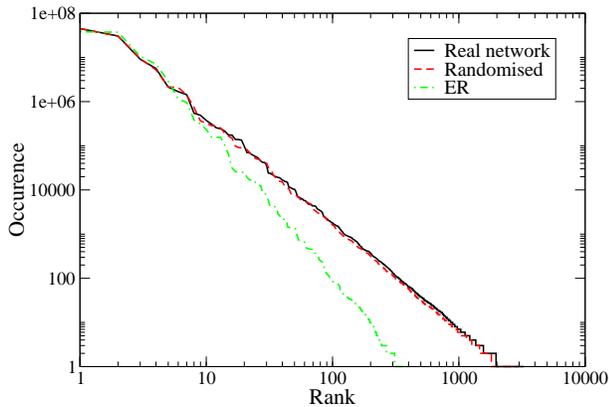}
\caption{Zipf plot for N=9 subgraphs of the giant component  of
  the E. coli network, for sample size $10^8$.  Also shown are the
  Zipf plots for the rewired network and for a Bernoulli or
  Erd{\"o}s-Renyi (ER) random graph with the same link probability and
  same sample size. Fixing the degree sequence almost exactly fixes
  the Zipf plot while the specific subgraphs exchange rank
  under rewiring.}
\label{fig:zipf}
\end{figure}

We first consider subgraphs with a fixed number of nodes and ask what
is the frequency of occurrence of different subgraphs.  For each $N>5$
a sample of $10^8$ subgraphs were obtained. The ensembles for $N=3$
and $N=4$ do not have enough subgraphs to obtain a smooth
distribution.  The labels $L$, $l_{1}$ and $l_{2}$ were used to
identify graphs, but loop counting was not included. The subgraphs
were then ranked in descending order of occurrence, and Zipf plots
were made~\cite{zipf}.

\begin{figure}[!tb]
\includegraphics[angle=0,width=8.0cm]{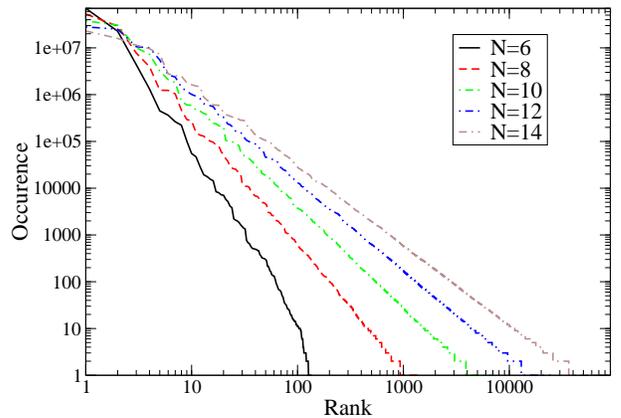}
\caption{Zipf plots obtained from the giant component of the E. coli
  network for subgraphs with varying numbers of nodes $N$.}
\label{fig:comparen}
\end{figure}

The Zipf plots all indicate power law behaviour.  Figure~\ref{fig:zipf}
shows a typical example. The distribution obtained from the GC
was compared to two different null cases. The first, denoted
``randomised'' in Fig.~\ref{fig:zipf}, is a rewired version of the
GC with the degree of each node fixed.  This was generated
by repeatedly choosing two links in the network at random and swapping their
endpoints, until mixing was achieved. As usual, mixing was evaluated
{\em a postiori}.  Swaps are disallowed if they create self-loops or produce a
pre-existing link.  The second null model is a random Erd{\"o}s-Renyi
(ER) network with the same link probability as the real network. For
the GC of the E. coli network, this link probability is
$p=695/{230 \choose 2} \approx 0.0264$, for the original network
$p=716/{270 \choose 2} \approx 0.0197$.  An ensemble of $10^8$ graphs
with the desired number of nodes was generated using a Bernoulli
process. In particular, a random number was placed on each pair of
distinct nodes to determine whether or not a link would be made.
This ensemble is denoted ``ER'' in the Zipf plots. As
demonstrated in Fig.~\ref{fig:zipf}, the Zipf plots of the real and
randomised networks are almost identical, but differ noticeably from
the ER network. This is true for all $N$, and for both the GC the entire E. coli network.

Figure~\ref{fig:comparen} shows Zipf plots for the GC with
varying subgraph sizes. It can be seen that all five sizes are
consistent with power law behaviour, although $N=6$ is less smooth than
the others because there are fewer subgraphs. The main difference
between the Zipf plots is that as $N$ increases the gradient becomes
shallower. Hence, it appears that the exponent is not universal with respect to $N$.

\begin{figure}[!tb]
\includegraphics[angle=0,width=8.0cm]{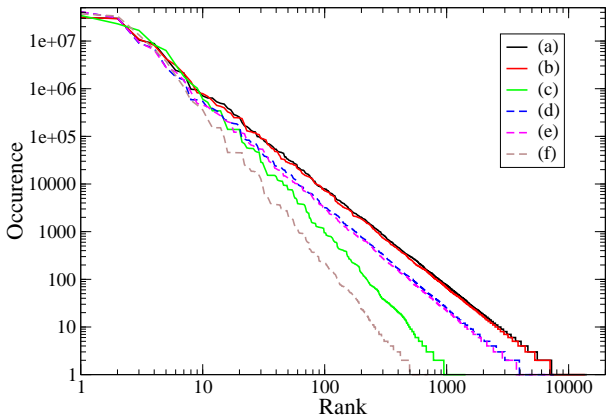}
\caption{Zipf plots for $N=11$ subgraphs in different networks: (a)
  real E. coli network, (b) rewired E. coli network, (c) ER network
  with same number of nodes and link probability as (a), (d) giant
  component of E. coli network, (e) rewired giant component, and (f)
  ER network with same number of nodes and link probability as (d).}
\label{fig:connect}
\end{figure}

Zipf plots for the original network and its GC are also
similar. As Fig.~\ref{fig:connect} shows, the plots for the real
network and the randomised network with identical degree sequence are
close in both cases.  The main difference between the GC
and the entire network is that in the latter case the distribution is
somewhat broader. However, the curves for the ER networks with
corresponding link probabilities show the same tendency, which
suggests that the difference in link probability may be the main
factor for this trend.

\subsection{Evidence for Motifs}

\begin{figure}[!tb]
\includegraphics[angle=0,width=8.0cm]{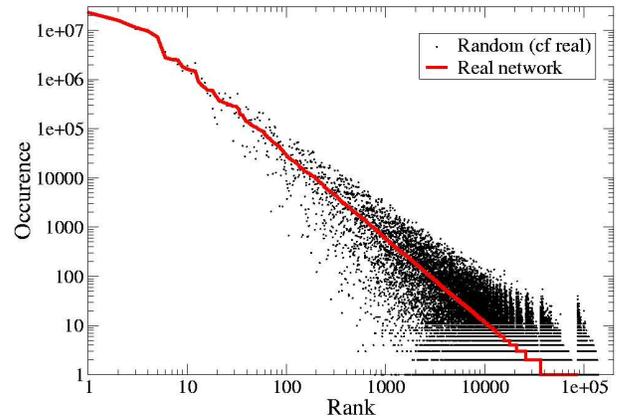}
\caption{Occurrences of $N=14$ subgraphs for the real (red line) and
  randomised (black points) networks. The subgraphs in the randomised
  network are placed along the $x$-axis in the same order as those in
  the real network to allow direct comparison between counts for each
  subgraph. Points significantly below the line represent motifs,
  while those significantly above represent anti-motifs.}
\label{fig:n14}
\end{figure}

Although the collection of subgraph counts are almost
identical for the real and randomised networks, the rank of individual
subgraphs within each census differs markedly.  The subgraphs of
the randomised network were arranged in the same order as those in the
real network to get the scatter plot shown in Fig.~\ref{fig:n14}. For
comparison, the Zipf plot for the real network is also shown as a
connected line. The vertical difference between each point and the
line indicates the difference in the number of occurrences of a
particular subgraph in the randomised network as compared to the
original one. Note that the rank of the subgraph in the original network gives
a unique tag to that subgraph. It can clearly be seen that the counts of certain
individual subgraphs vary by orders of magnitude between the two networks.

\begin{figure}[!tb]
\includegraphics[angle=0,width=8.0cm]{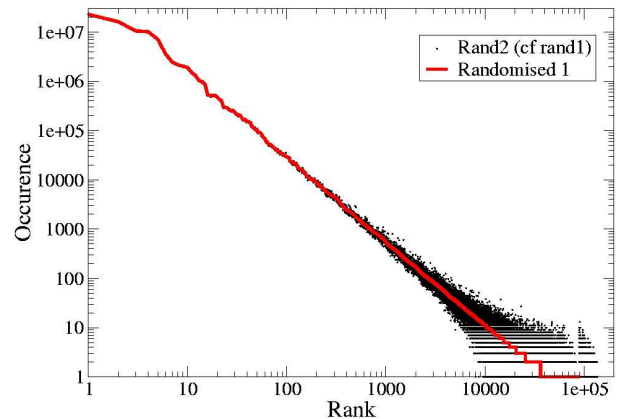}
\caption{This graph compares two randomised versions of the E. coli
  network in exactly the same way that the network and a randomised
  version of it were compared in Fig~\ref{fig:n14}.  The fluctuations,
  or scatter below and above the line in Fig.~\ref{fig:n14} are much
  larger, indicating a pattern of statistically significant deviations of
subgraph occurrences in the original network.}
\label{fig:n14ran}
\end{figure}

These large differences are not just a statistical artifact of the
rewiring process, as can be seen by re-doing the Fig.~\ref{fig:n14} for two
randomised networks with the same degree sequence as the E. coli
network. Now the subgraphs are ordered according to their occurrence
in the first randomised network.  Comparing Figs.~\ref{fig:n14} and
\ref{fig:n14ran}, note that the scatter  of points around the line
(particularly below the line) in the latter case is significantly less
than the former. This suggests the existence of ``motifs'' \cite{mot1,
  mot2, mot3}: particular subgraphs that are significantly
over-abundant  in the real network compared to its
ensemble of randomised networks.

To explore the issue of motifs further, subgraph counts from the real
network were compared to counts from several randomised networks. For
$N=3$ and $N=4$, we made an exhaustive enumeration of every subgraph.
This was done for the real network and one hundred different
randomised networks. The mean and standard deviation of the randomised
counts  were then computed, allowing a $Z$-score to
be calculated.  Fig.~\ref{fig:oldtable} shows the results for the
original network (with $Z$-scores for the GC in brackets).
The counts in the ER column are theoretical expectation values for an
ER network of the appropriate size and link probability.

\subsection{Linear Constraints Between Subgraphs}
\label{sec:linear}

For $N=3$ all the $Z$-scores all have the same magnitude.  This is a
direct consequence of the strict conservation of the degree sequence
in the rewiring procedure~\cite{wassermann_book}. Consider a particular swap between two
links. The only 3-node graphs that can possibly be affected are those
that contain at least one of the newly created or newly deleted links.
At least two of the three nodes must therefore be chosen from among
the four at the ends of the swapped links. The 4-node graph formed by
the swapping nodes themselves is always unchanged by all allowed swaps
(recall that it is not permitted to duplicate a pre-existing link).
Its 3-node subgraphs are therefore also unaffected.  The remaining
possibility involves 5-node graphs containing one extra node in
addition to the four swapping nodes. This extra node can have between
zero and four links connecting it to the four swapping nodes. It turns
out that there are only three pairs of 5-node graphs that can be
interchanged by link swapping.  In every case, the count of $N=3$
graphs with no links decreases by one, that of one-link graphs
increases by three, that of two-link graphs decreases by three and
that of three-link graphs increases by one (up to an overall sign).
This exact equality produces coincidence of the $Z$-scores.

The only remaining degree of freedom for the
deviation of the actual network from its randomised ensemble is a single signed number.  Its
value indicates a significant difference between the real network and
random networks with the same degree sequence, although it is
impossible to ascribe this significance to any one subgraph in
particular.  Note that for the empty 3-node graph, its statistical
under-abundance in the real network is due to the fact that the
variance of this number in the ensemble is tiny, because those changes
are slaved to a variable (the connected triangle) with small numbers.
The actual under-abundance of empty 3-node subgraphs is an unimportant
fraction of the overall number of those subgraphs.  This illustrates
the potential difficulties with assigning importance to individual subgraphs
based on their individual $Z$-score -- when the $Z$-scores must be correlated.

Conservation rules for subgraphs under rewiring was previously
observed by~\cite{sp} for 3-node subgraphs in directed networks, where
although there are thirteen different connected motifs, only seven
degrees of freedom are independent.  For undirected
$N$-node graphs, there are $N$ conservation laws corresponding to
moments of $\sum_i k_i^m$ with $m=0,1 \dots N-1$.  Hence for $N=3$
there is only one independent degree of freedom while for $N=4$ there
are seven.

\subsection{Motif Selection}

Ignoring the potential problems associated with attaching physical
importance to specific subgraphs with high individual $Z$-scores, we
find that for $N=4$ two graphs stand out as being particularly over-
or under-abundant. The square graph labelled $l_1=1679616$ is
over-represented, while the same graph with one edge missing
($l_1=6350400$) is under-represented. It is also interesting to note
that graphs with more (less) links tend towards over
(under)-abundance.  Overall the $Z$-scores are modestly lowered for
the GC, but the same overall trends emerge in both cases.
In particular, the same two subgraphs are readily identified as motif
and anti-motif.

For $N \ge 5$, an exhaustive scan of all subgraphs is
time-consuming, so uniform samples of $10^8$ subgraphs were used
instead.  Subgraphs do not need to be fully connected in order to be
useful for identifying motifs.
As for $N=3$ and $N=4$, the real network was compared to an ensemble
of randomised networks with the same degree sequence. Only 20 networks
were included, though, rather than 100.  Twenty was chosen as the
smallest number for which standard deviations and $Z$-scores are
reasonably stable. Checks show that when calculations are repeated,
the $Z$-scores obtained vary slightly, but the same graphs always
stand out as motifs.

The main difficulty is that too many subgraphs have high individual
$Z$-scores. This may be related to the correlations discussed above.
Ignoring previously mentioned caveats, we proceeded by using other
selection criteria to choose the most important. After some
experimentation the following $ad$ $ hoc$ rules were used to identify
motifs.  Two different samples were taken from the real network, and
$Z$-scores were computed comparing each of these to the same ensemble
of 20 randomised networks. A subgraph was identified as either a motif
(if it was connected) or containing a motif (if it was disconnect) if
$Z>10$ (or $Z<-10$ for anti-motifs) for both samples. Note that we
only consider connected pieces to be motifs even though the subgraphs
from which motifs are identified may be disconnected.  Requiring
$|Z|>10$ for two different samples largely eliminates statistical
oddities, which can otherwise occur for subgraphs with low counts. The
relatively high cut-off in $Z$ also helps ensure statistical
stability, as was also noted in \cite{samp}.  Even then, the number of
new motifs identified increases dramatically with $N$. To overcome
this problem, only subgraphs whose $Z$-scores were in the top fifty
for that value of $N$ were considered. Again, this had to be true for
both samples.  Motifs identified at a given $N$ tend to reappear as
connected components in disconnected graphs at higher $N$ -- see for
example the graph labelled 4096(1) for $N=3$ and $N=4$ in
Fig.~\ref{fig:oldtable}. The last condition was therefore that a new
motif has to replace an old one in the top fifty to make the grade.

Since including extra, unconnected nodes does not change the label of
a graph it easy to identify and eliminate previous motifs at each new
value of $N$.  Motifs with a given number of nodes are not always
discovered straight away; for example an $N=6$ motif may not meet the
condition $Z>10$ in the sample of $N=6$ subgraphs but show up much
more strongly (with one or two disconnected nodes) at $N=7$ or $N=8$.
This often means that subgraphs which only just fail the criteria at
one $N$ are positively identified at the next. This trend makes the
selection of motifs more robust against small changes in the rules
used to identify motifs. At some point, however, the number of genuine
new motifs found begins to account for a smaller and smaller
proportion of newly identified subgraphs. We also found that for $N>9$
a smaller proportion of sampled subgraphs had $|Z|>10$. Because of
these diminishing returns, the present search was stopped after
$N=10$.

There are several possible reasons for this loss of efficiency: one is
the finite size of the E.coli network, or 
another property of the network. It is also possible that the heuristic
may be starting to fail, recalling that the most accurate
version was not employed because of time constraints.  Wrongly
classifying a small percentage of nonisomorphic graphs as isomorphic
is unlikely to make much difference, but if the problem worsened,
genuine motifs could be swamped by other subgraphs which are more
common in the randomised networks. This potential difficulty does not
cast doubt on the motifs or antimotifs presented here, as none of them
fall into the categories of graphs that cause problems, which have
been thoroughly investigated for $N \le 8$. However, further
investigation might be appropriate before attempting to use this
method for much larger subgraphs.

The original network was considered first, then calculations were
repeated on the GC of the network the first few $N$. The
same motifs were identified for both networks, although the order in
which they were found varied slightly. We therefore conclude that the
technique is robust.

\subsection{Patterns in Motifs}

The motifs found are shown in Figs.~\ref{fig:motifs} (over-abundant)
and~\ref{fig:antimotifs} (under-abundant). Some striking patterns
appear.  First, many of the motifs have a bipartite structure where
the vertices can be divided into two sets such that no links exist
within either set, but many links exist between members of opposite
sets. Many graphs display a complete matching: each vertex is
connected to every member of the other set. Many more graphs have
almost complete matchings, missing just one or two links. Again, some
graphs are almost bipartite, with complete or almost complete
matchings between two sets of vertices, and just a few matchings
within each set. Some of these latter graphs may be seen as
interpolating between bipartite graphs and complete graphs, where
every vertex connected to every other vertex. Complete graphs at $N=4$
and $N=5$ are observed as motifs. All motifs have a high link:node
ratio. In fact, $L \ge N$ for all motifs. Finally, the
remaining motifs fall into one of the categories described above, with
the addition of one or two ``hanging'' links.

Antimotifs follow a completely different pattern.  They occur mostly
as trees or may contain at most a single loop (usually a triangle, but
there are two pentagons and one square) with long tails.  This is to
be contrasted with the bipartite structures of the over-abundant
subgraphs, which typically contain many loops. They also have fewer
links than motifs: either $L=N-1$ for pure chains or $L=N$, if there
is one loop). This difference in the link:node ratios is readily
apparent in Table~\ref{tab:twoway}. In fact, for a given $N$ no
overlap in $L$ values for motifs and antimotifs exists.

\begin{table}[!b]
  \caption{The number of motifs (bold) and antimotifs (italics) with a
    given number of nodes, $N$, and links, $L$.  The two classes are
    separated in this space, and do not overlap.}
\label{tab:twoway}
{\vskip 6pt}
\begin{ruledtabular}
\begin{tabular}{|cccccccccccc|}
\hline
 & \multicolumn{11}{|c|}{\textbf{L}} \\ \hline
{\bf N} & 3 & 4 & 5 & 6 & 7 & 8 & 9 & 10 & 11 & 12 & 13 \\ \hline
4 & {\em 1} & {\bf 1} & {\bf 1} & {\bf 1} & & & & & & & \\
5 & & {\em 2} & {\bf 1} & {\bf 1} & {\bf 3} & {\bf 2} & {\bf 1} & {\bf 1} & &
 & \\
6 & & & {\em 5} & {\em 4} & {\bf 4} & {\bf 4} & {\bf 5} & {\bf 4} & {\bf 3} &
 {\bf 2} & {\bf 1} \\
7 & & & & {\em 7} & {\em 10} & {\bf 2} & {\bf 5} & {\bf 5} & {\bf 4} &
 {\bf 3} & \\
8 & & & & & {\em 4} &{\em 4} & & & {\bf 1} & {\bf 1} & \\ \hline
\end{tabular}
\end{ruledtabular}
\end{table}

\section{Summary}
\label{sec:conclusions}

This paper addresses some of the problems associated with finding
extended structures in complex networks.  We propose a new heuristic
for graph isomorphism and validate its accuracy for classifying all
undirected subgraphs with $N$ up to 8. A version of the algorithm is
used, together with uniform sampling, to obtain statistical signatures
of the ensemble of $N$-node subgraphs in an E. coli protein
interaction network for subgraphs with $N$ up to 14. The distribution
of subgraph occurrences follows a power law and the Zipf plots do not change
significantly under rewiring.  Sampling all possible subgraphs for
various $N$ allows for the discovery of extended motifs.  Motifs are
considered to be individual, connected graphs that are vastly
over-represented in the network compared to a null model. They have
more edges, for a given number of nodes, than antimotifs and generally
display a bipartite structure or tend towards a complete graph.  In
contrast, antimotifs, are mostly trees or contain at most a single,
small loop.  The heuristic for graph isomorphism developed here can be
applied with minor changes to directed graphs.

\begin{acknowledgments}
  We thank Peter Grassberger for enlightening discussions about
  problems in sampling and significance estimation. M.~P. thanks Lee
  Smolin and the Perimeter Institute for their hospitality during the
  initiation of this research, and Stefan Boettcher for conversations
  about Zipf plots and motif discovery.
\end{acknowledgments}

\begin{widetext}
\clearpage

\begin{figure}[!tb]
\centering
\includegraphics[angle=0,width=18.0cm]{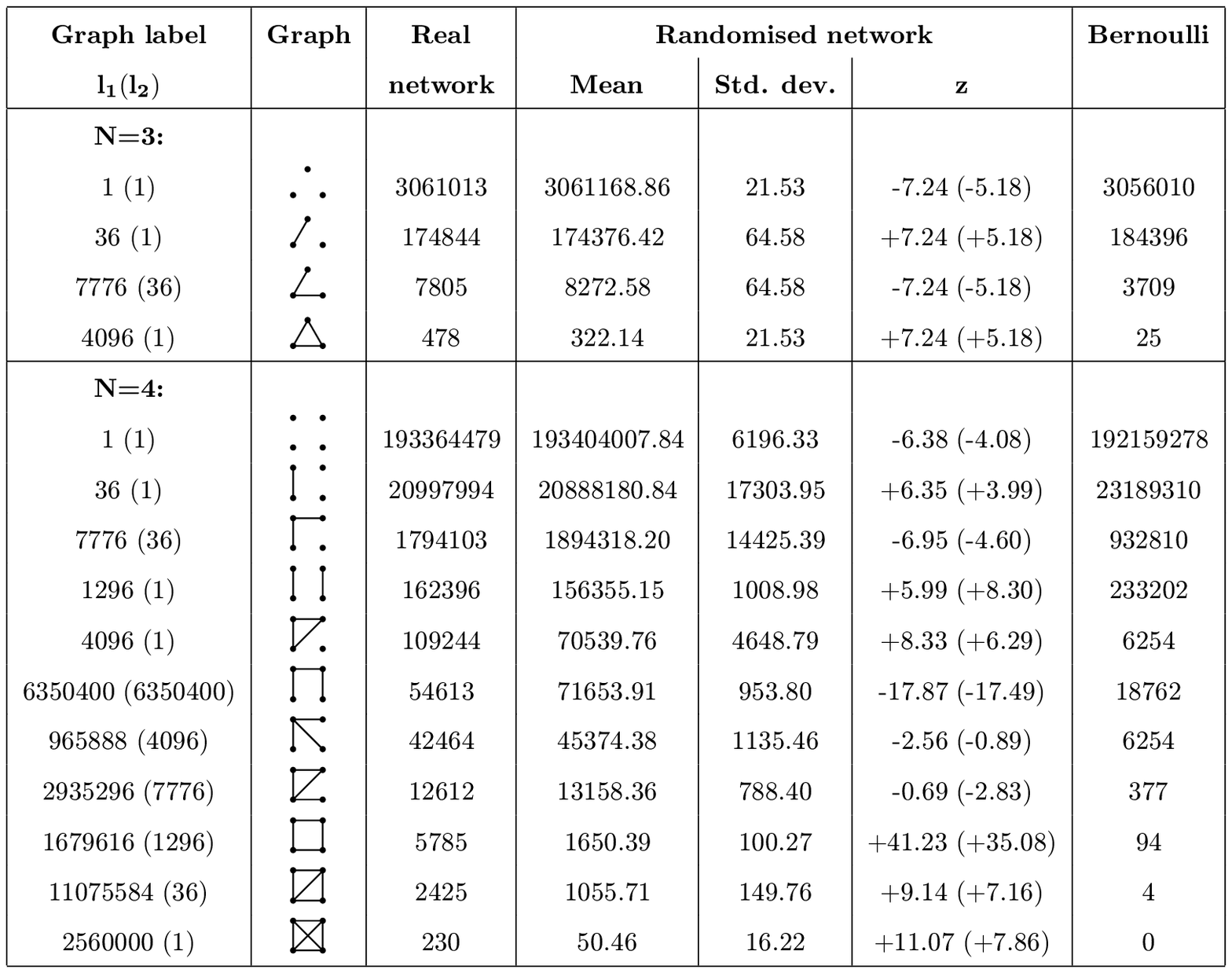}
\caption{Results for subgraphs with $N=3$ and $N=4$ nodes in the E.
  coli protein interaction network. The third column shows the counts
  obtained by exact enumeration for the real network, while columns
  4-6 show results obtained from exact enumeration of subgraphs in an
  ensemble of 100 networks with the same degree sequence. Standard
  deviations for the giant component are shown in brackets in column
  6. The last column shows theoretical expectation values for ER
  random graphs.}
\label{fig:oldtable}
\end{figure}

\clearpage

\begin{figure}[!tb]
\centering
\includegraphics[angle=0,width=17.0cm]{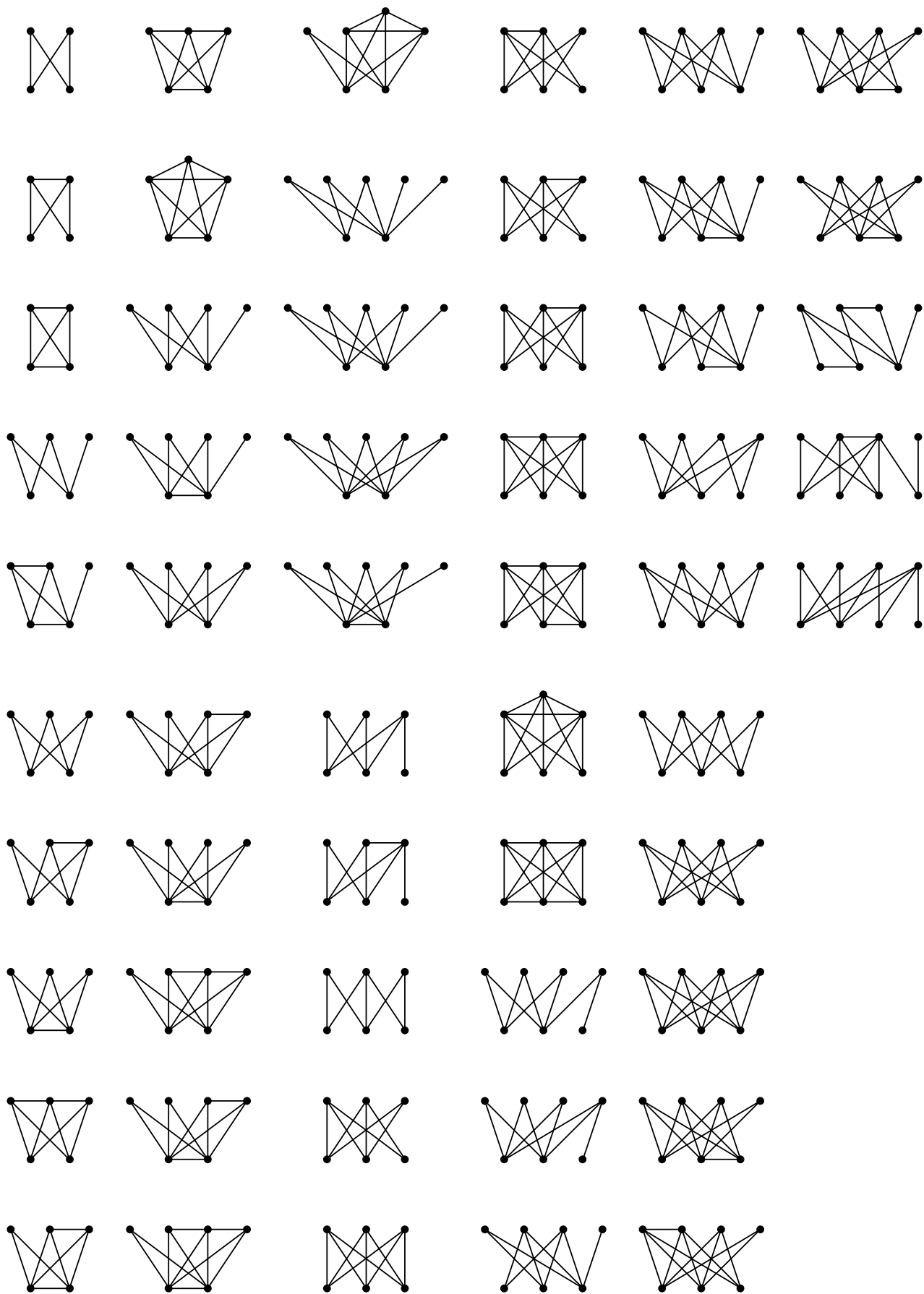}
\caption{Motifs (over-abundant subgraphs) of the E.coli protein
  interaction network.}
\label{fig:motifs}
\end{figure}

\clearpage

\begin{figure}[!tb]
\centering
\includegraphics[angle=0,width=18.0cm]{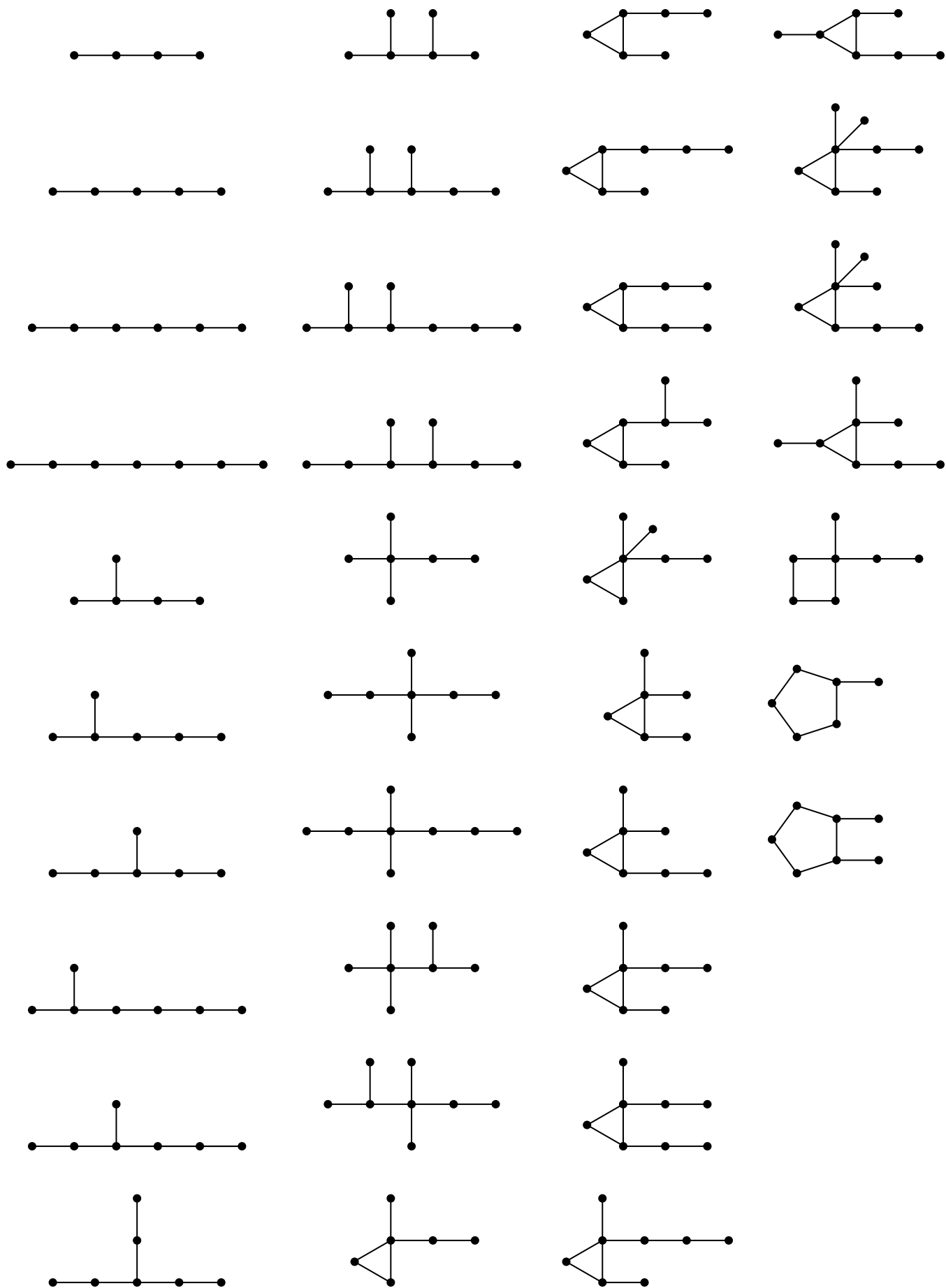}
\caption{Antimotifs (under-abundant subgraphs) of the E.coli protein
  interaction network.}
\label{fig:antimotifs}

\end{figure}
\clearpage

\end{widetext}

\label{lastpage}


\begin{thebibliography}{15}
\bibitem{foot1} It is not known to fail for larger
  graphs, but exhaustive testing is not practical, nor does it appear
  to be necessary for our purposes.
\bibitem{bases} Different bases could be used in the definition of
  $X_i$, for example $X_i = \sum_j b_1^{x-x_{ij}} b_2^{k_j}$, where
  $b_1$ and $b_2$ could in principle be different constants. Several
  possibilities were tested, but $b_1 = b_2 = 2$ was found to be most
  efficient.
\bibitem{zipf}{The Zipf plot is related to the probability
    distribution of occurrence. If a subgraph $i$ has rank $R_{i}$ and
    occurs $n_{i}$ times, then $R_{i}$ is the number of graphs that
    occur $n_{i}$ times or more.  Hence $R_{i} = \int_{n_i}^{\infty}
    P(n)$, where $P(n)$ is the probability for a subgraph to occur $n$
    times. For a power law distribution $P(n) \propto n^{-\gamma}$ so
    $R_i \propto n_i^{1-\gamma}$, or $n_i \propto
    R_i^{\frac{1}{1-\gamma}}$.}
\end{thebibliography}
\end{document}